\documentclass[conference]{IEEEtran}

\usepackage[utf8]{inputenc}

\usepackage{amssymb}
\usepackage{epigraph}
\usepackage{graphicx}
\usepackage[hyphens]{url}
\usepackage{hyperref}
\usepackage{float}

\usepackage{pdfpages}
\usepackage{tlatex}

\usepackage{todonotes}

\newcommand\tlaplus{TLA$^+$}
\newcommand\logv{\texttt{log}}
\newcommand\readIndex{\texttt{readIndex}}
\newcommand\commitIndex{\texttt{commitIndex}}
\newcommand\epoch{\texttt{epoch}}

\newcommand\Keys{\texttt{Keys}}
\newcommand\Values{\texttt{Values}}
\newcommand\NoValue{\texttt{NoValue}}

\newcommand\VersionBound{\texttt{VersionBound}}
\newcommand\StalenessBound{\texttt{StalenessBound}}

\newcommand\WriteConsistencyLevel{\texttt{WriteConsistencyLevel}}

\newcommand\StrongConsistency{\texttt{StrongConsistency}}

\newcommand\SessionConsistency{\texttt{SessionConsistency}}

\newfloat{listing}{htbp}{loe}
\floatname{listing}{Listing}

\begin{document}

\title{Understanding Inconsistency in Azure Cosmos DB with TLA+}

\author{
	\IEEEauthorblockN{A. Finn Hackett}
	\IEEEauthorblockA{University of British Columbia}
	Vancouver, Canada \\
	fhackett@cs.ubc.ca
	\and
	\IEEEauthorblockN{Joshua Rowe}
	\IEEEauthorblockA{Microsoft}
	Redmond, USA \\
	joshua.rowe@microsoft.com
	\and
	\IEEEauthorblockN{Markus Alexander Kuppe}
	\IEEEauthorblockA{Microsoft Research}
	Redmond, USA \\
	makuppe@microsoft.com
}

\maketitle

\begin{abstract}
	Beyond implementation correctness of a distributed system, it is equally important to understand exactly what users should expect to see from that system.
	Even if the system itself works as designed, insufficient understanding of its user-visible semantics can cause bugs in its dependencies.
	By focusing a formal specification effort on precisely defining the expected user-facing behaviors of the Azure Cosmos DB service at Microsoft, we were able to write a formal specification of the database that was significantly smaller and conceptually simpler than any other specification of Cosmos DB, while representing a wider range of valid user-observable behaviors than existing more detailed specifications.
	Many of the additional behaviors we documented were previously poorly understood outside of the Cosmos DB development team, even informally, leading to data consistency errors in Microsoft products that depend on it.
	Using this model, we were able to raise two key issues in Cosmos DB's public-facing documentation, which have since been addressed.
    We were also able to offer a fundamental solution to a previous high-impact outage within another Azure service that depends on Cosmos DB.
\end{abstract}


\section{Introduction}

Consistency guarantees for distributed databases are notoriously hard to understand.
Not only can distributed systems inherently behave in unexpected and counter-intuitive ways due to internal concurrency and failures, but they can also lull their users into a false sense of functional correctness: most of the time, users of a distributed database will witness a much simpler and more consistent set of behaviors than what is actually possible.
Only timeouts, fail-overs, or other rare events will expose the true set of behaviors a user might witness~\cite{cloudincident2019}.
Testing for these scenarios is difficult at best: reproducing them reliably requires controlling complex concurrency factors, latency variations, and network behaviors.
Even just producing usable documentation for developers is fundamentally challenging~\cite{apilearning2011,programmingquestions2008,apidocumentationfail2015}, and explaining these subtle consistency issues via documentation comes as an additional burden to distributed system developers and technical writers alike.
Formal methods have long been applied to the design of distributed systems, including in industry~\cite{amazons3-2015,shardstore2021,everest2017,amazons2n2018}, but these are years-long high-effort projects that focus on implementation correctness, not explaining the system to users.
Rather than focus on this difficult task, we address a simpler and more fundamental question: ignoring the implementation, what kind of behavior \emph{should} a client be able to witness while interacting with a service?

We use \tlaplus\ to answer this simpler question for Cosmos DB, a planet-scale key-value store.
In practice, Cosmos DB offers a rich interface featuring multiple query APIs, and has complex operational behaviors involving georeplication and partitioning of data.
As our focus is on data consistency from a client perspective, we model only the core read and write operations underlying the system's semantics relating to their 5 configurable consistency levels.
We show that this minimal client-focused specification of a large-scale service offers important design- and documentation-level insights, while keeping buy-in cost low.

We document the 2 person-month development process of our specification, which consists of iterative prototyping using the public documentation~\cite{cosmosdocs2022}, feedback from author 2, a Cosmos DB developer, and the specification and model checking of a collection of formal properties based on our understanding.
Aside from the specification itself, we discuss a pair of key issues it helped us discover within Cosmos DB's documentation, and how both have since been addressed.
We also use our specification to explain the previously-unclear root cause of a 28-day high-priority outage within Microsoft Azure.


We describe the following results: (1) a concise (390 LOC) client-focused specification of Cosmos DB, a large-scale distributed system; (2) a pair of key documentation bugs we found by developing our specification --- statements in Cosmos DB's public documentation~\cite{cosmosdocs2022} that have now been corrected; and (3) using our specification, a novel and concise mechanized explanation of a high-severity Cosmos DB-related outage within Azure that took 28 days to identify and mitigate.

Beyond our work so far, we expect our model to be useful in future design work as Cosmos DB's implementation evolves, aided by its ability to precisely and abstractly state a client's expectations of system behavior.
Services depending on Cosmos DB may also benefit from incorporating our work into \tlaplus\ specifications of their own processes, in which case our work may be used to prevent future outages similar to the one we describe in this paper.

\section{Background}

Our work uses the \tlaplus\ specification language~\cite{lamport2002specifying}, which can be used to describe state machines using set-theoretic constructs and temporal logic.
Models written in \tlaplus\ have no direct correspondence to implementations, with users focusing instead on analyzing design decisions and verifying model-level correctness properties.
This philosophy allows specification writers to leave out irrelevant details and focus on expressing a specification's core semantics as simply as possible.

In addition to plain \tlaplus, model developers can also write models in PlusCal~\cite{pluscal2009}, a high-level imperative language that is more like contemporary programming languages.

It is possible to check model properties using the explicit-state model checker TLC~\cite{tlatoolbox2019,tlc1999}, the symbolic model checker Apalache~\cite{apalache2019}, and the manual proof assistant TLAPS~\cite{tlaps2012}.
In this work, we relied on the TLC model checker to analyze our specification.

As well as model checking temporal properties, it is also possible to express and check refinements~\cite{refinement1988} in \tlaplus.
A refinement proves that one specification implements another -- meaning that one specification exhibits every behavior that another specification exhibits, given an appropriate translation between the two specifications' state spaces.
We use this technique to show that our new specification produces a superset of the behaviors produced by existing \tlaplus\ specifications of Cosmos DB.

\section{A Simple Model of Cosmos DB}

To fully illustrate our claim to simplicity, this section describes our full formalization of Cosmos DB's semantics in a few pages, including most of the core \tlaplus\ definitions in-text.
Our modeling process was based on iterative discussion with author 2, a principal engineer working on the Cosmos DB implementation.
We followed existing user-facing documentation, asked for feedback, learned more about the realities of Cosmos DB's design, and incorporated that new knowledge into our specification.
We repeated this feedback loop until we found no more corrections, when our model began to predict counter-intuitive but possible behaviors of the real system.
See Subsection \ref{scn:notable-anomalies} for in-depth analysis of such behaviors.
See \url{https://github.com/tlaplus/azure-cosmos-tla/tree/master/simple-model} for the full \tlaplus definitions.

\subsection{Consistency Levels}

\begin{figure}
    \centering
    \includegraphics[width=.4\textwidth]{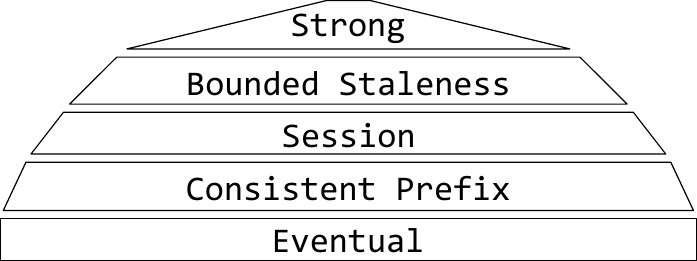}
    \caption{Hierarchy of consistency levels in Cosmos DB, with strongest at the top and weakest guarantees at the bottom.}
    \label{fig:consistency-pyramid}
\end{figure}

Cosmos DB offers 5 consistency levels that affect read and write behavior.
A system administrator must configure all writes to follow a single consistency level per Cosmos DB deployment.
Read operations may either match the configured write consistency level or weaken it according to the hierarchy defined in Figure \ref{fig:consistency-pyramid}.
We discuss high-level prose descriptions of these consistency levels, which we complement with precise \tlaplus\ descriptions later on.

\textbf{Strong consistency.} Reads and writes are linearizable~\cite{linearizability1990}, as long as the operation does not fail.
See Section \ref{scn:notable-anomalies} for possible behaviors given failures.

\textbf{Bounded staleness.} Writes older than a given time bound are durable and consistently readable, whereas writes younger than the given bound are not.
The time bound is defined two ways: a bound in wall-clock time, and a maximum bound on the number of eventually-consistent writes allowed at once.
If the bounds are in danger of being exceeded, additional writes will be refused for replication to catch up.
For modeling simplicity, we ignore the wall-clock temporal aspect of this mode's semantics, and consider only operation count.

\textbf{Session consistency.} Reads and writes are tagged with \emph{session tokens}.
Operations with the same session token are linearizable relative to one another, but no guarantees are provided across different session tokens.
Session consistency writes are not guaranteed durable, and tokens may be invalidated by data loss.

\textbf{Consistent prefix.} Reads are monotonic: a client may only read newer values than it has already read.
Section \ref{scn:consistent-eventual-equiv} describes why we find this is ultimately equivalent to eventual consistency.

\textbf{Eventual consistency.} This level offers no ordering guarantees, but does provide a notion of eventual convergence over an arbitrary period of time.

\subsection{Data Definitions}

Our model of Cosmos DB is defined to have 4 state variables, and allows them to evolve over time via some simple actions.
Each variable relates to a specific aspect of the system being modeled.

\textbf{log.} The \logv\ is a sequence of writes, represented as key-value pairs.
For example, $[\texttt{key} \mapsto \texttt{k1},\ \texttt{value} \mapsto \texttt{v1}]$ pairs key $\texttt{k1}$ with value $\texttt{v1}$.
The sequence lists all writes that are stored anywhere in Cosmos DB's implementation, irrespective of replication or durability.

\textbf{readIndex.} The \readIndex\ marks either a position in the \logv\ or $0$.
For any element of the \logv, if its index is less than or equal to \readIndex, then it is replicated universally across the current Cosmos DB deployment.
Representing eventually-complete propagation of writes, the log prefix defined by \readIndex\ behaves identically to a single key-value store.

\textbf{commitIndex.} The ``commit index" marks a position in the \logv\ or $0$.
For any element of the \logv, if its index is less than or equal to \commitIndex, then it is replicated at a global majority of replicas, and is therefore durable due to consensus.
It follows from this definition that $\readIndex~\leq~\commitIndex$ must always hold.

\textbf{epoch.} The \epoch\ is a monotonically increasing counter of fail-overs.
If \epoch\ remains constant, fail-overs behavior such as data loss may not be observed.
If it increases, some data loss may be observed at the point of increase.

The specification has six constants: \Keys\ and \Values, which are the sets of keys and values respectively.
These sets can be redefined based on the use case --- they can be generalized to infinite sets like ``all strings", or restricted to a small finite set of constant values in order to simplify model checking.
\NoValue\ is a constant indicating the absence of a value.
\VersionBound\ and \StalenessBound\ are natural numbers that affect when writes are allowed.
\WriteConsistencyLevel\ represents the currently configured consistency level for write operations, one of the 5 consistency levels.

We chose to base our specification on a sequential log because Cosmos DB, like any consensus-based system, determines a total order in which clients should consider their requests to have occurred.
This is why many parts of our specification, including several state variables, identify writes by log index.

Building on these definitions, we can express our first two fundamental actions. 

\begin{figure}[h]
\begin{tla}
IncreaseReadIndexAndOrCommitIndex ==
    /\ commitIndex' \in commitIndex..Len(log)
    /\ readIndex' \in readIndex..commitIndex'
    /\ UNCHANGED <<log, epoch>>
\end{tla}
\begin{tlatex}
\@x{ IncreaseReadIndexAndOrCommitIndex \.{\defeq}}%
 \@x{\@s{16.4} \.{\land} commitIndex \.{'} \.{\in} commitIndex \.{\dotdot} Len
 ( log )}%
 \@x{\@s{16.4} \.{\land} readIndex \.{'} \.{\in} readIndex \.{\dotdot}
 commitIndex \.{'}}%
\@x{\@s{16.4} \.{\land} {\UNCHANGED} {\langle} log ,\, epoch {\rangle}}%
\end{tlatex}
\end{figure}

\textit{IncreaseReadIndexAndOrCommitIndex} models the concept of data replication, that is, \readIndex\ and/or \commitIndex\ advancing.
\readIndex\ and \commitIndex\ may non-deterministically gain new values between \readIndex\ and $\commitIndex'$, and \commitIndex\ and $\textbf{Len}(\logv)$, respectively.
Neither \logv\ nor \epoch\ may change.
This ensures that both values may only grow, that they never point beyond end of the log, and that $\readIndex \leq \commitIndex$ remains true.

\begin{figure}[h]
\begin{tla}
TruncateLog ==
    \E i \in (commitIndex+1)..Len(log) :
        /\ log' = SubSeq(log, 1, i - 1)
        /\ epoch' = epoch + 1
        /\ UNCHANGED <<readIndex, commitIndex>>
\end{tla}
\begin{tlatex}
\@x{ TruncateLog \.{\defeq}}%
 \@x{\@s{16.4} \E\, i \.{\in} ( commitIndex \.{+} 1 ) \.{\dotdot} Len ( log )
 \.{:}}%
\@x{\@s{27.72} \.{\land} log \.{'} \.{=} SubSeq ( log ,\, 1 ,\, i \.{-} 1 )}%
\@x{\@s{27.72} \.{\land} epoch \.{'} \.{=} epoch \.{+} 1}%
 \@x{\@s{27.72} \.{\land} {\UNCHANGED} {\langle} readIndex ,\, commitIndex
 {\rangle}}%
\end{tlatex}
\end{figure}

\textit{TruncateLog} models the concept of data loss: if there exists any index $i$ such that $\commitIndex < i$, then \logv\ may be truncated non-deterministically such that its new length is $i - 1$.
In-progress operations may watch for changes in \epoch's value to detect and respond to failures, meaning \epoch\ acts as a failure detector.

Because these actions may happen non-deterministically, any combination of replication and fail-over may occur at any time, interleaved with other actions.
A short sequence of such actions can represent a complex series of implementation-level possibilities.

\subsection{Write Operations}

In Cosmos DB, write operations are not atomic.
They may sometimes appear atomic under certain configurations\footnote{For instance, a client performing only strongly consistent reads and strongly consistent writes will never witness an in-progress write.
Weaker consistency levels do not provide any such guarantees, however.
See Section \ref{scn:notable-anomalies} for specific examples.}, but their underlying structure needs to be broken down into multiple steps.

As a consequence of writes' multi-step nature, we need to record the state of in-progress writes.
For portability, we don't require any particular state retention mechanism, as the specifics might vary depending on how our core specification is used.
Instead, we break up the two conceptual stages of a Cosmos DB write into re-usable parts that we describe individually.

\subsubsection{Beginning a Write Operation}

\begin{figure}[h]
\begin{tla}
WritesAccepted ==
    /\ Len(logv) - readIndex < VersionBound
    /\ ((WriteConsistencyLevel = BoundedStaleness) =>
        Len(logv) - commitIndex < StalenessBound)
\end{tla}
\begin{tlatex}
\@x{ WritesAccepted \.{\defeq}}%
\@x{\@s{16.4} \.{\land} Len ( logv ) \.{-} readIndex \.{<} VersionBound}%
 \@x{\@s{16.4} \.{\land} ( ( WriteConsistencyLevel \.{=} BoundedStaleness )
 \.{\implies}}%
\@x{\@s{31.39} Len ( logv ) \.{-} commitIndex \.{<} StalenessBound )}%
\end{tlatex}
\end{figure}

\textit{WritesAccepted} determines whether a write may be attempted at all.
It constrains writes based on two factors: \VersionBound and \StalenessBound.
\VersionBound\ is a global limit on how many partially-replicated writes may exist in a Cosmos DB instance at any one time.
\StalenessBound\ is a global limit on how many non-durable writes may exist in a Cosmos DB instance at any one time, used to enforce bounded staleness consistency.

\begin{figure}[h]
\begin{tla}
WriteInit(key, value) ==
    log' = Append(log, [key |-> key, value |-> value])
\end{tla}
\begin{tlatex}
\@x{ WriteInit ( key ,\, value ) \.{\defeq}}%
 \@x{\@s{16.4} log \.{'} \.{=} Append ( log ,\, [ key \.{\mapsto} key ,\,
 value \.{\mapsto} value ] )}%
\end{tlatex}
\end{figure}

\textit{WriteInit} defines the initial stage of any permitted write operation, appending a new key-value pair to the log.
This means that at least one replica now holds the new key-value pair.
The lack of distinction between incomplete and complete writes is intentional here: Cosmos DB replicas unconditionally begin serving writes as soon as they accept them.
The Cosmos DB client libraries are the ones that enforce Cosmos DB's read semantics, and they may perform multiple read requests against multiple replicas until they get a consistent answer that can be returned to an end-user.
Cosmos DB replicas require no additional logic restricting which writes should be visible to which read requests.

\begin{figure}[h]
\begin{tla}
WriteInitToken ==
    [epoch |-> epoch, checkpoint |-> Len(log) + 1]
\end{tla}
\begin{tlatex}
\@x{ WriteInitToken \.{\defeq}}%
 \@x{\@s{16.4} [ epoch \.{\mapsto} epoch ,\, checkpoint \.{\mapsto} Len ( log
 ) \.{+} 1 ]}%
\end{tlatex}
\end{figure}

\textit{WriteInitToken} defines a unique identifier, or token, with which we can keep track of a write's progress.
This token is structurally identical to a session token, the data used to identify a client's session at session consistency.
Note that in practice, these tokens represent the flow of a request from client to server and back.
We use this abstraction to concisely summarize an otherwise complex mix of network semantics and client-server interaction.

We have model-checked a uniqueness property for all session tokens given up to 6 writes and any one failure event.

\subsubsection{Completing a Write Operation}

Once it has begun, a write operation may complete \emph{at any time that it is allowed to}.
An in-progress write may also non-deterministically fail at any time, due to timeouts, spurious network failures, and so forth.

\begin{figure}[h]
\begin{tla}
WriteCanSucceed(token) ==
    /\ SessionTokenIsValid(token)
    /\ (WriteConsistencyLevel = StrongConsistency =>
            token.epoch = epoch)
\end{tla}
\begin{tlatex}
\@x{ WriteCanSucceed ( token ) \.{\defeq}}%
\@x{\@s{16.4} \.{\land} SessionTokenIsValid ( token )}%
 \@x{\@s{16.4} \.{\land} ( WriteConsistencyLevel \.{=} StrongConsistency
 \.{\implies}}%
\@x{\@s{47.79} token . epoch \.{=} epoch )}%
\end{tlatex}
\end{figure}

Given a token identifying an in-progress write, \textit{WriteCanSucceed} defines when the write is allowed to succeed.
There are 3 conditions for success.

First, a write may succeed if its token is valid, that is, \\$\textit{SessionTokenIsValid}(\texttt{token})$ is true.
This will be the case if \\$\texttt{token}.\texttt{checkpoint} \leq \textbf{Len}(\logv)$, and $\texttt{token}.\texttt{epoch} = \epoch$.
	
Second, writes must still be in the log.
If data loss occurred and the written data is gone, success cannot be claimed.
This condition also accounts for replacement, where log entries are lost then written again with the same index.
Since data loss always increments \epoch, rejecting writes from a different \epoch\ cleanly disallows writes interrupted by data loss events.
	
Lastly, if \WriteConsistencyLevel\ is set to \StrongConsistency, then $\texttt{token}.\texttt{checkpoint}$ must be less than or equal to \commitIndex.
By the semantics of \commitIndex, this requirement means that ``all successful strongly consistent writes must be durable".

\subsection{Read Operations}

We define read semantics for Cosmos DB as stateless, read-only operators that describe the set of allowed read results for any given read request.
We define the read operation for each consistency level separately, but we use a common underlying definition called \textit{GeneralRead} to avoid duplication.

\begin{figure}[h]
\begin{tla}
GeneralRead(key, index, allowDirty) ==
    LET maxCandidateIndices == { i \in DOMAIN log :
            /\ log[i].key = key
            /\ i <= index }
        allIndices == { i \in DOMAIN log :
            /\ allowDirty
            /\ log[i].key = key
            /\ i > index }
    IN  { [logIndex |-> i, value |-> log[i].value]
            : i \in allIndices \cup (
            IF   maxCandidateIndices # {}
            THEN {Max(maxCandidateIndices)}
            ELSE {}) } \cup 
        (IF   maxCandidateIndices = {}
        THEN {NotFoundReadResult}
        ELSE {})
\end{tla}
\begin{tlatex}
\@x{ GeneralRead ( key ,\, index ,\, allowDirty ) \.{\defeq}}%
 \@x{\@s{16.4} \.{\LET} maxCandidateIndices \.{\defeq} \{ i \.{\in} {\DOMAIN}
 log \.{:}}%
\@x{\@s{51.43} \.{\land} log [ i ] . key \.{=} key}%
\@x{\@s{51.43} \.{\land} i \.{\leq} index \}}%
\@x{\@s{39.13} allIndices \.{\defeq} \{ i \.{\in} {\DOMAIN} log \.{:}}%
\@x{\@s{55.53} \.{\land} allowDirty}%
\@x{\@s{55.53} \.{\land} log [ i ] . key \.{=} key}%
\@x{\@s{55.53} \.{\land} i \.{>} index \}}%
 \@x{\@s{16.4} \.{\IN}\@s{4.09} \{ [ logIndex \.{\mapsto} i ,\, value
 \.{\mapsto} log [ i ] . value ]}%
\@x{\@s{51.01} \.{:} i \.{\in} allIndices \.{\cup} (}%
\@x{\@s{51.01} {\IF}\@s{8.2} maxCandidateIndices \.{\neq} \{ \}}%
\@x{\@s{51.01} \.{\THEN} \{ Max ( maxCandidateIndices ) \}}%
\@x{\@s{51.01} \.{\ELSE} \{ \} ) \} \.{\cup}}%
\@x{\@s{39.13} ( {\IF}\@s{8.2} maxCandidateIndices \.{=} \{ \}}%
\@x{\@s{39.13} \.{\THEN} \{ NotFoundReadResult \}}%
\@x{\@s{39.13} \.{\ELSE} \{ \} )}%
\end{tlatex}
\end{figure}

\textit{GeneralRead} takes 3 parameters: \texttt{key}, whose value is being read; \texttt{index}, a log index indicating the reader's ``point of view" in the log; and \texttt{allowDirty}, which determines whether the read operation should have exactly one result, or non-deterministically many.
All members of the resulting set will be pairs of \texttt{logIndex} and \texttt{value}, which are resulting value and its log index, respectively.
\texttt{logIndex} allows read results to be totally ordered, which is useful for both verifying correctness properties, and for correctly describing session tokens.

\texttt{index} defines a prefix of the log, selecting all indices $i \leq \texttt{index}$.
Within this prefix, \textit{GeneralRead} will always include the latest mapping from \texttt{key} to some value.
If there is no such mapping, \textit{GeneralRead} returns the marker value \textit{NotFoundReadResult}.
Additionally, if \texttt{allowDirty} is true, then \textit{GeneralRead} will also include values bound to \texttt{key} in log entries with index $i > \texttt{index}$.
This models non-deterministic reads: it allows reading writes that are not durable, still in progress, or simply arbitrarily more recent than \texttt{index}.

Note that each of the following read operations are only valid for compatible values of \WriteConsistencyLevel.
Figure \ref{fig:consistency-pyramid} illustrates the intended hierarchy of consistency levels. 

\subsubsection{Strongly Consistent Reads}

\begin{figure}[h]
\begin{tla}
StrongConsistencyRead(key) ==
    GeneralRead(key, commitIndex, FALSE)
\end{tla}
\begin{tlatex}
\@x{ StrongConsistencyRead ( key ) \.{\defeq}}%
\@x{\@s{16.4} GeneralRead ( key ,\, commitIndex ,\, {\FALSE} )}%
\end{tlatex}
\end{figure}

Strongly consistent reads for any given key follow \commitIndex, and return one single consistent value in all cases.
Aligning these reads with \commitIndex\ means that only durable writes may be read.

\subsubsection{Bounded Staleness Reads}

\begin{figure}[h]
\begin{tla}
BoundedStalenessRead(key) ==
    GeneralRead(key, commitIndex, TRUE)
\end{tla}
\begin{tlatex}
\@x{ BoundedStalenessRead ( key ) \.{\defeq}}%
\@x{\@s{16.4} GeneralRead ( key ,\, commitIndex ,\, {\TRUE} )}%
\end{tlatex}
\end{figure}

Bounded staleness reads also follow \commitIndex.
Unlike strongly consistent reads, bounded staleness reads may see arbitrary information beyond \commitIndex.
The span of log entries between \commitIndex\ and $\textbf{Len}(\logv)$ represents the non-durable reads allowed, which may be arbitrarily witnessed in addition to durable data before \commitIndex.

\subsubsection{Session Consistent Reads}

\begin{figure}[h]
\begin{tla}
SessionConsistencyRead(token, key) ==
    IF   \/ epoch = token.epoch
         \/ token = NoSessionToken
    THEN LET sessionIndex == Max({token.checkpoint, 
                                  readIndex})
         IN  GeneralRead(key, sessionIndex, TRUE)
    ELSE {})
\end{tla}
\begin{tlatex}
\@x{ SessionConsistencyRead ( token ,\, key ) \.{\defeq}}%
\@x{\@s{16.4} {\IF}\@s{8.2} \.{\lor} epoch \.{=} token . epoch}%
\@x{\@s{35.19} \.{\lor} token\@s{0.25} \.{=} NoSessionToken}%
 \@x{\@s{16.4} \.{\THEN} \.{\LET} sessionIndex \.{\defeq} Max ( \{ token .
 checkpoint ,\,}%
\@x{\@s{166.73} readIndex \} )}%
\@x{\@s{44.69} \.{\IN} GeneralRead ( key ,\, sessionIndex ,\, {\TRUE} )}%
\@x{\@s{16.4} \.{\ELSE} \{ \} )}%
\end{tlatex}
\end{figure}

Session consistent reads operate using a session token which defines a position in the log to read from: a \texttt{checkpoint}, and the \epoch\ from which the token originates.

The first check made during a session consistency read is whether the session token is from the current \epoch.
If the epochs differ, and the session token isn't the placeholder value \textit{NoSessionToken}, then no reads are permitted.
Session consistency offers no durability guarantees: if data loss occurs, it becomes impossible to guarantee that writes referenced by a session token remain intact.
Not all session tokens will be invalidated on every data loss event in practice, but we have yet to find a need for modeling the invalidation of only some session tokens.

After checking the \epoch, the \texttt{checkpoint} is combined with \readIndex.
Since the \readIndex\ indicates the log prefix that has been replicated to every single replica in the current Cosmos DB deployment, it would be unsound to have a \texttt{sessionIndex} smaller than \readIndex.

We set \texttt{allowDirty} to \texttt{TRUE}, meaning that a session consistent read may arbitrarily read log entries beyond its session token.
This possibility represents clients' ability to non-deterministically witness the effects of other concurrent sessions.

Note that the ``empty" value, \textit{NoSessionToken}, corresponds to $[\texttt{epoch} \mapsto 0, \texttt{checkpoint} \mapsto 0]$.
Its \epoch\ of $0$ makes it incomparable to other session tokens, and its checkpoint of $0$ places no constraint on the outcome of a session consistency read.

\begin{figure}[h]
\begin{tla}
UpdateTokenFromRead(origToken, read) == [
    epoch |-> epoch,
    checkpoint |-> Max({origToken.checkpoint,
                        read.logIndex})
]
\end{tla}
\begin{tlatex}
\@x{ UpdateTokenFromRead ( origToken ,\, read ) \.{\defeq} [}%
\@x{\@s{16.4} epoch \.{\mapsto} epoch ,\,}%
\@x{\@s{16.4} checkpoint \.{\mapsto} Max ( \{ origToken . checkpoint ,\,}%
\@x{\@s{106.43} read . logIndex \} )}%
\@x{ ]}%
\end{tlatex}
\end{figure}

Once a read is performed with a given token, a client must update its session token.
This is done with \textit{UpdateTokenFromRead}, which combines the log index from a read result with the checkpoint of an existing session token.
This combination monotonically increases a client's session token, ensuring that each client may only witness increasingly recent information.

\subsubsection{Consistent Prefix and Eventual Consistency Reads}

\begin{figure}[h]
\begin{tla}
EventualConsistencyRead(key) ==
    GeneralRead(key, readIndex, TRUE)
\end{tla}
\begin{tlatex}
\@x{ EventualConsistencyRead ( key ) \.{\defeq}}%
\@x{\@s{16.4} GeneralRead ( key ,\, readIndex ,\, {\TRUE} )}%
\end{tlatex}
\end{figure}

Consistent prefix and eventual consistency being known equivalent, as discussed in Section \ref{scn:consistent-eventual-equiv}, they have identical definitions.
Their behavior is minimally constrained, requiring only that values overwritten at or before \readIndex\ cannot be read.

\subsection{Validation}

To validate that our specification exhibits behaviors of which Cosmos DB's implementation is capable, and in order to ensure that we cover as wide a variety of these behaviors as possible, we have leveraged a combination of model checking correctness properties, model checking our specification's relationship with comparable specifications via refinement, and manual expert review of behaviors implied by our model.
This subsection focuses on the properties we checked, while particularly interesting specific behaviors will be discussed alongside our results in Section \ref{scn:results}.

\subsubsection{Correctness Properties}

The correctness properties we check are a collection of the ones listed in Cosmos DB's external documentation~\cite{cosmosdocs2022}, properties derived from existing \tlaplus\ specifications of Cosmos DB~\cite{cosmostla2018} (which are also referenced as authoritative by Cosmos DB's documentation), and properties inherent to our particular specification's design.
To aid in our verification process, we extend our base behavior specification with an auxiliary \texttt{writeHistory} state variable.
\texttt{writeHistory} provides a history of all attempted writes, including which key, which value, a write token indicating at which epoch and log index the write began, and a state that will transition at most once from \texttt{WriteInitState} to either \texttt{WriteSucceededState} or \texttt{WriteFailedState}.

Using this extended specification, we verify a total of 10 liveness properties and 14 safety properties across the 4 distinct data consistency levels offered by Cosmos DB, excluding basic type safety invariants.
Our verification process is based on model checking, using a combination of exhaustive state space exploration of logs up to length 6, and depth-first random simulation of execution traces exploring up to 100 steps.

\begin{figure}[h]
\begin{tla}
PointsValid ==
    [][/\ readIndex <= commitIndex
       /\ readIndex <= readIndex'
       /\ commitIndex <= commitIndex']_vars
\end{tla}
\begin{tlatex}
\@x{ PointsValid \.{\defeq}}%
\@x{\@s{16.4} {\Box} [ \.{\land} readIndex \.{\leq} commitIndex}%
\@x{\@s{26.65} \.{\land} readIndex \.{\leq} readIndex \.{'}}%
\@x{\@s{26.65} \.{\land} commitIndex \.{\leq} commitIndex \.{'} ]_{ vars}}%
\end{tlatex}
\end{figure}

For example, \texttt{PointsValid} defines the relationship between \readIndex\ and \commitIndex: \readIndex\ cannot be beyond \commitIndex, and they must increase monotonically.
For the sake of concision, we describe the other properties via prose summary.
The full set is available alongside our complete specification at \url{https://github.com/tlaplus/azure-cosmos-tla/tree/master/simple-model}.

\textbf{Read your writes.} For strong consistency and session consistency with the same token, after any write, only the written value or some later write may be read.

\textbf{Read after write.} It is similar to \emph{read your writes} at this level of abstraction, changing from global to client perspective. 

\textbf{Monotonic reads.} For strong consistency and session consistency with the same token, reads may only make visible later writes, and will never return older data than they already have.

\textbf{Bounded staleness.} Bounded staleness consistency should never accept more than \StalenessBound\ uncommitted writes at once.

\textbf{Session token lifetime.} For any arbitrary session token that is valid, it will either remain valid or become invalid, in which case it will never become valid again.

\textbf{readIndex as lower bound}. No reads may return values that have been overwritten by other operations within the log prefix defined by \readIndex.

\textbf{Write completion.} All writes eventually complete, either with success or failure.

\subsubsection{Linearizability}

The strongest consistency property offered by Cosmos DB is the linearizability~\cite{linearizability1990,tlalinearizability-lorin} of write operations at the \StrongConsistency\ consistency level. 
Linearizability means that, for any operation on some concurrent object (here, a key in Cosmos DB), we can choose a point in time between the beginning and end of that operation at which it has atomically occurred.
For our simple model, that point is when \commitIndex\ is incremented.
Due to our non-atomic modeling of writes, this point will occur at some unspecified point in between the beginning and end of a successful write, whenever \textit{IncreaseReadIndexAndOrCommitIndex} takes place.
We wrote a refinement specification \texttt{CosmosDBLinearizability}, verifying that every behavior of our Cosmos DB specification with strong consistency reads and writes corresponds to the same series of atomic reads and writes applied to a \tlaplus\ function.

\subsubsection{Refining Existing Specifications}

Cosmos DB already has publicly available \tlaplus\ models for some of its behavior~\cite{cosmostla2018}, so we used refinement to verify our new specification does not disagree with existing specifications.

We found that our work offers a superset of previously-modeled behavior, despite the old specification including concepts we do not explicitly deal with, like synchronization between replicas in Azure regions.
Our model's behavior is specifically a strict superset of the old one's, because we noted that the existing specifications made no attempt at modeling data loss or relaxed reads.

\subsubsection{Refining Read Consistency Levels}

It is strongly hinted in Cosmos DB's public documentation~\cite{cosmosdocs2022} that different consistency levels represent a hierarchy of possible behaviors, with stronger consistency guarantees forming subsets of weaker consistency guarantees.
We used our specification to investigate this property, and determined under what conditions the implication made by Cosmos DB's public documentation holds.

We found that, \emph{for the same configured write consistency}, different consistency reads form behavioral subsets directly matching the documented hierarchy illustrated in Figure \ref{fig:consistency-pyramid}.
Keeping the write consistency level constant, each stronger read consistency allows a subset of the behavior of each weaker read consistency.
Note that we consider all possible session token choices together for session consistency.

Counter-intuitively, this relationship does not hold when comparing write consistency levels.
Consider that strong consistency allows more non-durable writes than bounded staleness, because bounded staleness fundamentally relies on throttling writes to preserve its semantics, whereas strong consistency does not.
See Section \ref{scn:bounded-staleness-weaker} for discussion.

\section{Results} \label{scn:results}

\begin{table}
    \caption{Read and write strategies at different consistency levels, taken from public Cosmos DB documentation~\cite{cosmosdocs2022}.}
    \label{tbl:consistency-levels}
    
    \centering
	\begin{tabular}{l | p{1.9cm} | l}
		\textbf{Consistency Level} & \textbf{Quorum Reads} & \textbf{Quorum Writes} \\
		\hline
		Strong & Local Minority & Global Majority \\
		Bounded Staleness & Local Minority & Local Majority \\
		Session & Single Replica (session token) & Local Majority \\
		Consistent Prefix & Single Replica & Local Majority \\
		Eventual & Single Replica & Local Majority
	\end{tabular}
\end{table}

Beyond our specification itself, we showcase two key issues it helped us raise with Cosmos DB's documentation, both of which have been addressed.
We also present the previously-unclear root cause of a 28-day high-priority outage within Azure, alongside a collection of other properties of Cosmos DB made explicit by our project.
Table \ref{tbl:consistency-levels} lists the semantics we use when providing replica- and region-level example scenarios.
At the implementation level, read and write operations at different consistency levels require different degrees of replication or consensus, which are listed in this table.

\subsection{Consistent Prefix and Eventual Consistency Behave Equivalently}
\label{scn:consistent-eventual-equiv}

During our work, we have discovered that eventual consistency and consistent prefix in Cosmos DB behave identically from the point of view of a client performing individual reads and writes.
Since eventual consistency is the least constraining option, we will focus on whether consistent prefix could behave in any way that is distinguishable from it.
For context, consider the original description of consistent prefix consistency below, which has now been rewritten by the Azure documentation team in response to our findings.

\epigraph{
    In consistent prefix option, updates that are returned contain some prefix of all the updates, with no gaps. Consistent prefix consistency level guarantees that reads never see out-of-order writes.
    
    If writes were performed in the order $\left<\texttt{A,B,C}\right>$, then a client sees either $\left<\texttt{A,A,B}\right>$, or $\left<\texttt{A,B,C}\right>$, but never out-of-order permutations like $\left<\texttt{A,C}\right>$ or $\left<\texttt{B,A,C}\right>$. Consistent Prefix provides write latencies, availability, and read throughput comparable to that of eventual consistency, but also provides the order guarantees that suit the needs of scenarios where order is important.
}{Azure Cosmos DB Documentation on Consistent Prefix~\cite{cosmosdocs2022}}

Looking at the examples in the above excerpt, the documentation claimed that neither $\left<\texttt{A,C}\right>$ nor $\left<\texttt{B,A,C}\right>$ should be observable by clients.
We assume the scenario described involves some implicit key $k$ and values $\left<\texttt{A,B,C}\right>$ written to key $k$ in sequence alongside 3 concurrent read operations, all under consistent prefix.
In that case, our model of Cosmos DB allows both sequences of reads that the documentation claims are forbidden.

The first sequence, $\left<\texttt{A,C}\right>$, is possible because the concurrent interleaving $\left<\texttt{write(A)},\right.$ \texttt{read(A)}, \texttt{write(B)}, \texttt{write(C)}, $\left.\texttt{read(C)}\right>$ should naturally be possible, even if all operations were globally atomic.
We are not sure why this counter-example was claimed to be invalid.

\begin{figure}
    \centering
    \includegraphics[width=.4\textwidth]{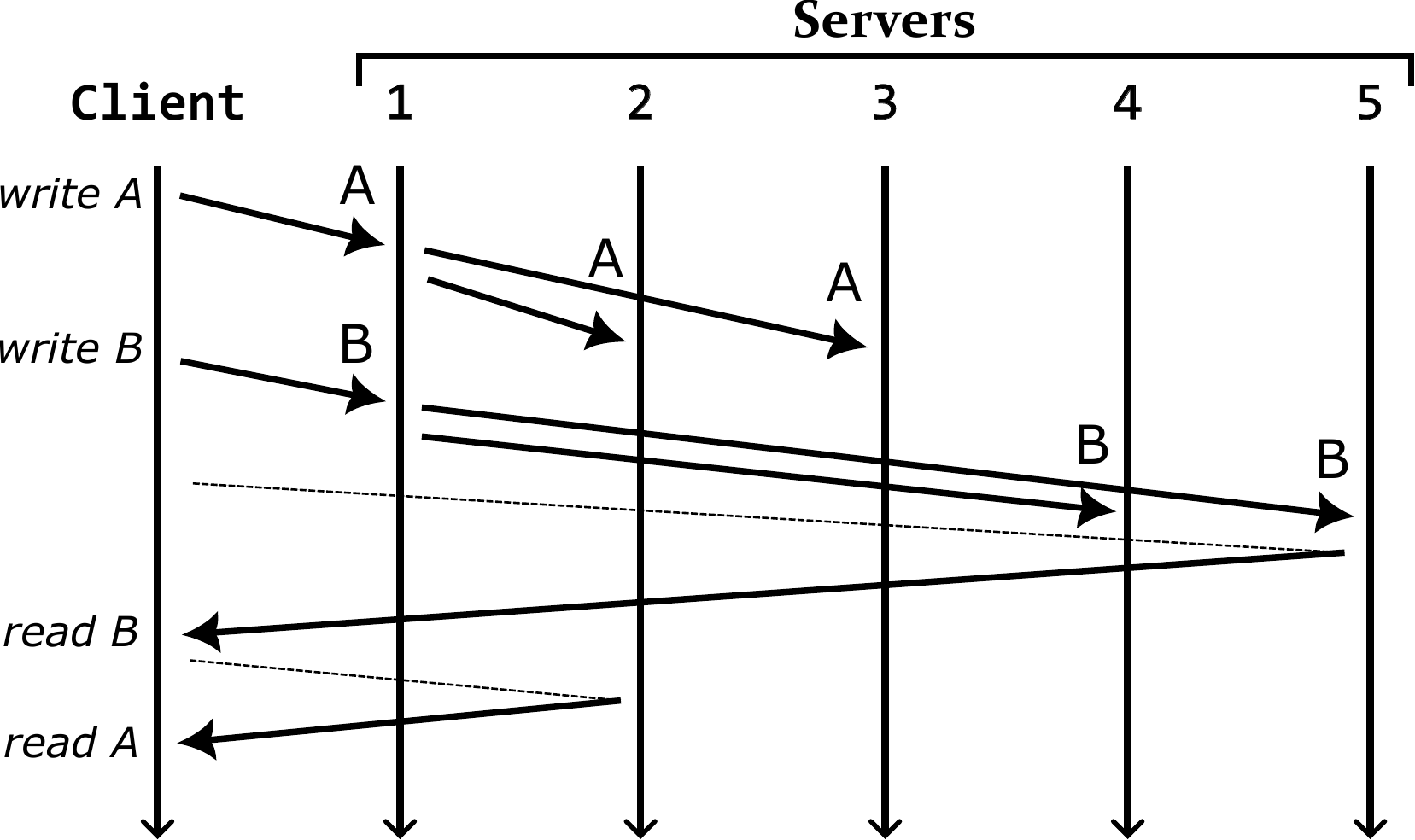}
    \caption{Interaction diagram of a possible scenario for producing read pattern $\left<\texttt{B,A}\right>$ under consistent prefix with a cluster of 5 servers.}
    \label{fig:counterexample-1}
\end{figure}

The second sequence, $\left<\texttt{B,A,C}\right>$, is a more complex case.
We have confirmed that read operations in Cosmos DB are load balanced to potentially any replica, and that any replica will immediately serve any data that is replicated to it.
Following the information from Table \ref{tbl:consistency-levels}, we know that consistent prefix read operations will go to only one replica, and that consistent prefix write operations will be considered successful once data has been committed by a local majority of replicas in a single region.
Figure \ref{fig:counterexample-1} illustrates a possible scenario with one client and 5 replicas that will produce the read pattern $\left<\texttt{B,A}\right>$~\footnote{We omit \texttt{C} from our example, as additionally writing then reading \texttt{C} after seeing $\left<\texttt{B,A}\right>$ is intuitive, strongly consistent behavior.}, while following all known implementation-level semantics of consistent prefix consistency.

First, assuming some arbitrary single key $k$, the client writes values \texttt{A} and \texttt{B} to local majorities.
There are 5 replicas, so 3 servers must commit each write.
The first write goes to replicas $1,2,3$, and the second write goes to replicas $1,4,5$.
Then, the client performs two reads in quick succession.
Due to arbitrary load balancer behavior, the first read is served by replica $5$, and the second read is served by replica $2$.
Each replica serves its latest local copy of the data bound to key $k$, which, due to how local majorities were chosen during the earlier writes, and assuming no replication has time to take place, produces the sequence of reads $\left<\texttt{B,A}\right>$.

Together, these two counter-examples negate the only documented difference between consistent prefix and eventual consistency for atomic writes to the same key.

\subsection{Regions Do Not Affect Safety Guarantees}

Building on the idea that consistent prefix and eventual consistency behaviors are identical, we arrive at a second question regarding Cosmos DB's public documentation: why is data consistency so strongly dependent on how regions are configured?
To illustrate, Cosmos DB's consistency documentation~\cite{cosmosdocs2022} contains 13 bullet points across 3 sections indicating consistency expectations that depend on the region in which a client is interacting with Cosmos DB.
12 of those bullet points list either consistent prefix or eventual consistency as the expected behavior, which we found to be equivalent.

Given how many of these bullet points be argued redundant according to our specification, we gave thought to whether they could all be removed.
Making the documentation simpler in this way would be a net positive to potential readers who seek to understand Cosmos DB's consistency guarantees.

The 13th bullet point that lists a consistency level other than eventual consistency or consistent prefix applies to bounded staleness, when bounded staleness reads go to the same region as writes.
That bullet point claims that, under those conditions, bounded staleness offers guarantees identical to strong consistency.
Given that both reads and writes under bounded staleness perform region-local consensus, we can understand why this case would often be equivalent to strong consistency in practice: within the same region, it would be impossible for a client to see any out-of-order artifacts.
The missing condition is durability: Cosmos DB supports write region fail-over, whereby the write region can be changed if the original write region has become unavailable.
In that case, the new write region might not have replicated all of the data in the original write region, or might lag behind other regions that are still available, allowing both data loss and stale reads.
This would not be the case for strong consistency, which requires global consensus during writes, meaning that changing write regions would not create any client-visible inconsistencies.

When our issues were addressed, it was confirmed that for atomic single reads and writes, our arguments are valid.
These bullet points remain due to an additional detail that is out of scope for our specification: transactions.
Cosmos DB supports optimistic concurrency via a transaction engine layered on top of the raw reads and writes our specification supports, which acts differently under consistent prefix and eventual consistency.
Formally specifying this new information, as we have done for the original, may be an interesting direction for future work.

\subsection{Investigating a High-Impact Production Outage}
\label{scn:sev1}

Our work was motivated by a past production outage, which was highlighted in an ongoing study of cloud incidents at Microsoft\footnote{Note to reviewers: we will add the reference once it becomes available.}.
The outage, found after 26 days in production by a customer, impacted thousands of subscriptions over 28 days, causing significantly increased error rates for certain kinds of resource allocation calls in Azure.

We have used our work to model the semantics underlying the outage.
As a result, we have been able to identify the previously-unidentified safety issue underlying the outage.
We presented our analysis to the author of the original outage postmortem, and they confirmed that our explanation made sense within the context of their work.

\subsubsection{Outage Postmortem and Investigation}

\begin{figure}
    \centering
    \includegraphics[width=.4\textwidth]{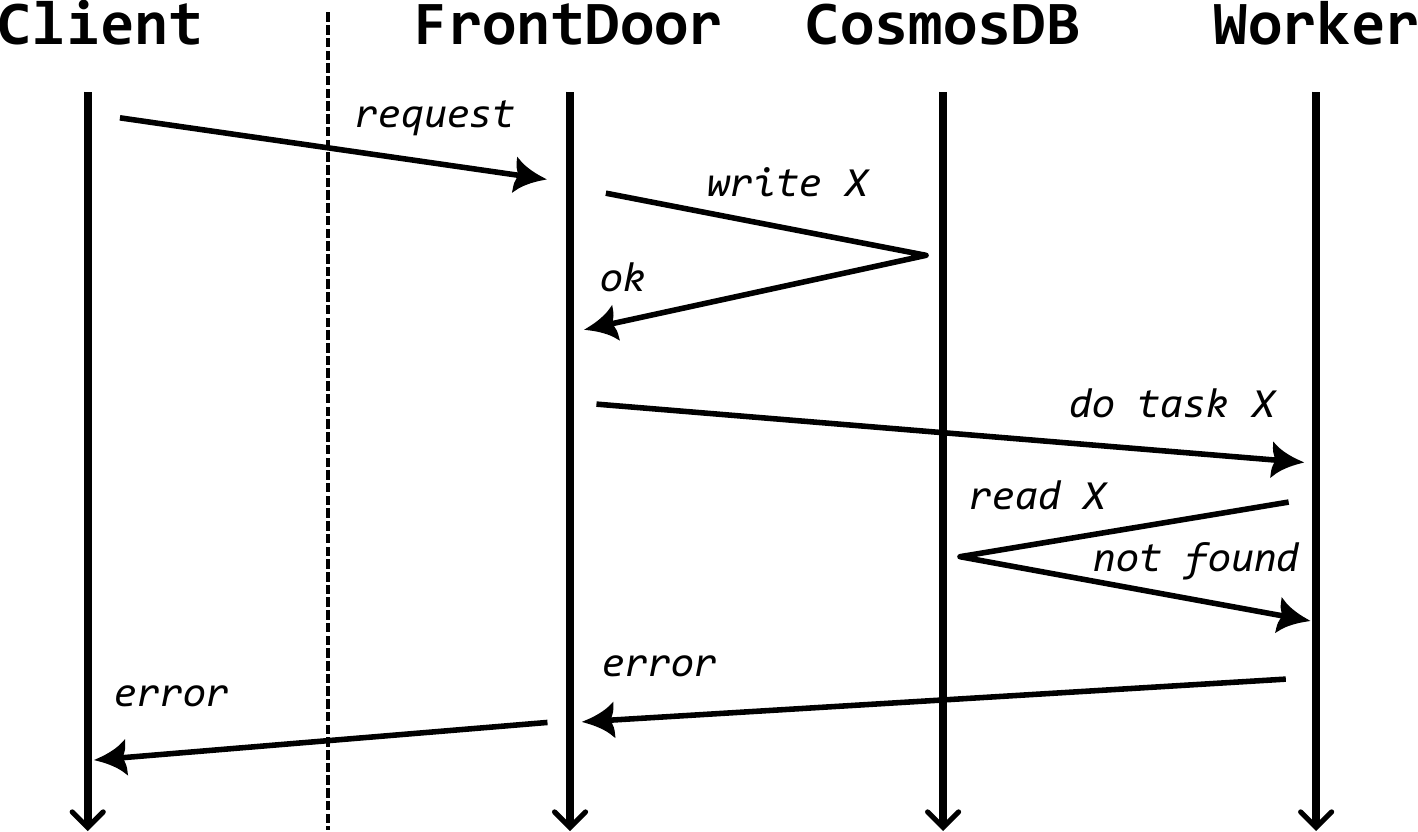}
    \caption{Interaction diagram of the error underlying a high-impact outage within Azure Cloud.}
    \label{fig:sev1-A}
\end{figure}

Figure \ref{fig:sev1-A} illustrates the underlying structure of the outage, as reported in the postmortem.
It is already clear that a data consistency issue is occurring: the \textit{FrontDoor} server makes a complete, successful write to Cosmos DB, and then the \textit{Worker} server tries to read that write and fails.
The reported explanation for this issue was based on regions and latency.
\textit{FrontDoor} was performing writes in one cloud region, and \textit{Worker} was reading those writes in another region.
Prior to the change, these writes and reads were occurring in the same region, and errors were not happening at a noticeable rate.
The change in routing lead to a change in latency, which caused \textit{Worker} to read out of date information from Cosmos DB.

From our analysis, while the original postmortem's comments were correctly diagnosing the change in latency, some correctness-critical factors were not discussed.
We found that Cosmos DB was configured for session consistency, and we confirmed that each server (\textit{FrontDoor}, \textit{Worker}) was working with unconfigured, arbitrary session tokens.
From the definition of session consistency, without sharing session tokens, the two servers were only guaranteed eventually consistent reads.
So, the semantic problem had always existed, but it was only exposed in practice by a change in region configuration.

\subsubsection{Modeling the Outage}

\begin{listing}
\raggedright
\begin{pcal}
variables serviceBus = <<>>;

process (frontdoor = "frontdoor")
variables frontdoorToken;
{
    frontdoorWriteTaskDataInit:
    assert CosmosDB!WritesAccepted;
    await CosmosDB!WriteInit("taskKey", "taskValue");
    frontdoorToken := CosmosDB!WriteInitToken;
    frontdoorWriteTaskDataCommit:
    await CosmosDB!WriteCanSucceed(frontdoorToken);
    serviceBus := <<"taskKey">>;
}

process (worker = "worker")
variables workerToken, workerValue;
{
    workerBeginTask:
    await serviceBus # <<>>;
    with(taskKey = Head(serviceBus),
    read \in CosmosDB!SessionConsistencyRead(
    CosmosDB!NoSessionToken, taskKey)) {
        serviceBus := Tail(serviceBus);
        workerToken :=
        CosmosDB!UpdateTokenFromRead(
        CosmosDB!NoSessionToken, read);
        workerValue := read.value;
    }
}
\end{pcal}
\begin{tlatex}
\@x{ {\p@variables} serviceBus \.{=} {\langle} {\rangle} {\p@semicolon}}%
\@pvspace{8.0pt}%
\@x{ {\p@process} {\p@lparen} frontdoor \.{=}\@w{frontdoor} {\p@rparen}}%
\@x{ {\p@variables} frontdoorToken {\p@semicolon}}%
\@x{ {\p@lbrace}}%
\@x{ frontdoorWriteTaskDataInit\@s{.5}\textrm{:}\@s{3}}%
\@x{\@s{16.4} {\p@assert} CosmosDB {\bang} WritesAccepted {\p@semicolon}}%
\@x{\@s{16.4} {\p@await} CosmosDB {\bang} WriteInit (\@w{taskKey}
    ,\,\@w{taskValue} ) {\p@semicolon}}%
\@x{\@s{16.4} frontdoorToken \.{:=} CosmosDB {\bang} WriteInitToken
    {\p@semicolon}}%
\@x{ frontdoorWriteTaskDataCommit\@s{.5}\textrm{:}\@s{3}}%
\@x{\@s{16.4} {\p@await} CosmosDB {\bang} WriteCanSucceed ( frontdoorToken )
    {\p@semicolon}}%
\@x{\@s{16.4} serviceBus \.{:=} {\langle}\@w{taskKey} {\rangle}
    {\p@semicolon}}%
\@x{ {\p@rbrace}}%
\@pvspace{8.0pt}%
\@x{ {\p@process} {\p@lparen} worker \.{=}\@w{worker} {\p@rparen}}%
\@x{ {\p@variables} workerToken ,\, workerValue {\p@semicolon}}%
\@x{ {\p@lbrace}}%
\@x{ workerBeginTask\@s{.5}\textrm{:}\@s{3}}%
\@x{\@s{16.4} {\p@await} serviceBus \.{\neq} {\langle} {\rangle}
    {\p@semicolon}}%
\@x{\@s{16.4} {\p@with} {\p@lparen} taskKey \.{=} Head ( serviceBus ) ,\,}%
\@x{\@s{45.70} read \.{\in} CosmosDB {\bang} SessionConsistencyRead (}%
\@x{\@s{45.70} CosmosDB {\bang} NoSessionToken ,\, taskKey ) {\p@rparen}
    {\p@lbrace}}%
\@x{\@s{39.03} serviceBus \.{:=} Tail ( serviceBus ) {\p@semicolon}}%
\@x{\@s{39.03} workerToken \.{:=}}%
\@x{\@s{39.03} CosmosDB {\bang} UpdateTokenFromRead (}%
\@x{\@s{39.03} CosmosDB {\bang} NoSessionToken ,\, read ) {\p@semicolon}}%
\@x{\@s{39.03} workerValue \.{:=} read . value {\p@semicolon}}%
\@x{\@s{16.4} {\p@rbrace}}%
\@x{ {\p@rbrace}}%
\end{tlatex}

\caption{A PlusCal model of the behavior underlying the events in Figure \ref{fig:sev1-A}.}
\label{lst:sev1-pluscal}
\end{listing}

Our \tlaplus\ model allows us to verify the abstract scenario from Figure \ref{fig:sev1-A}, and check our understanding against our specification of Cosmos DB.
In Listing \ref{lst:sev1-pluscal}, we use PlusCal to model this scenario.
We chose PlusCal in order to demonstrate a more implementation-like mode of interaction with our Cosmos DB model, which may be more familiar to implementation developers.
We include the majority of our PlusCal model of this scenario, with minor edits and omissions for presentation.

Similarly to Figure \ref{fig:sev1-A}, in Listing \ref{lst:sev1-pluscal} we have two processes called \texttt{frontdoor} and \texttt{worker}, which match 2 of the 4 processes in the figure.
We omit the client process by starting our model at the point where \texttt{frontdoor} begins to handle the client request.
We do not explicitly specify a process for Cosmos DB, since that is taken from our core specification.
The configuration that sets \WriteConsistencyLevel\ to \SessionConsistency\ is also omitted.

Cosmetic differences aside, the underlying series of actions is the same as in Figure \ref{fig:sev1-A}: the \texttt{frontdoor} writes some value (here, \texttt{X} is \texttt{"taskValue"}).
The write occurs in two steps: one to begin the write, and one to await the write's success.
Assuming the write succeeds, \texttt{frontdoor} writes \texttt{"taskKey"} to a service bus (modeled here as a global sequence), requesting that the worker perform some task named \texttt{"taskKey"} (\texttt{X} in the figure).
To perform the task, \texttt{worker} must read the task data from Cosmos DB.
It does this using a session consistency read with a null session token, storing the value it reads.

\subsubsection{Counter-Example}

Based on the original issue's data consistency expectations, we can formulate an expected property for our model in temporal logic: $\lozenge\ \texttt{workerValue} = \texttt{"workerValue"}$.
That is, eventually the state variable \texttt{workerValue} will hold the value \texttt{"workerValue"} written by \texttt{frontdoor}.
Model-checking our property generates a counter-example.
In that counter-example, \readIndex\ and \commitIndex\ were both at $0$, meaning no replication had taken place, and that unconstrained session-consistency reads could go to replicas that did not have our single write of \texttt{"taskValue"}.
Using a small amount of PlusCal alongside our model of Cosmos DB, we were able to accurately recreate the semantic issue underlying a high-impact outage.

\begin{figure}
    \centering
    \includegraphics[width=.4\textwidth]{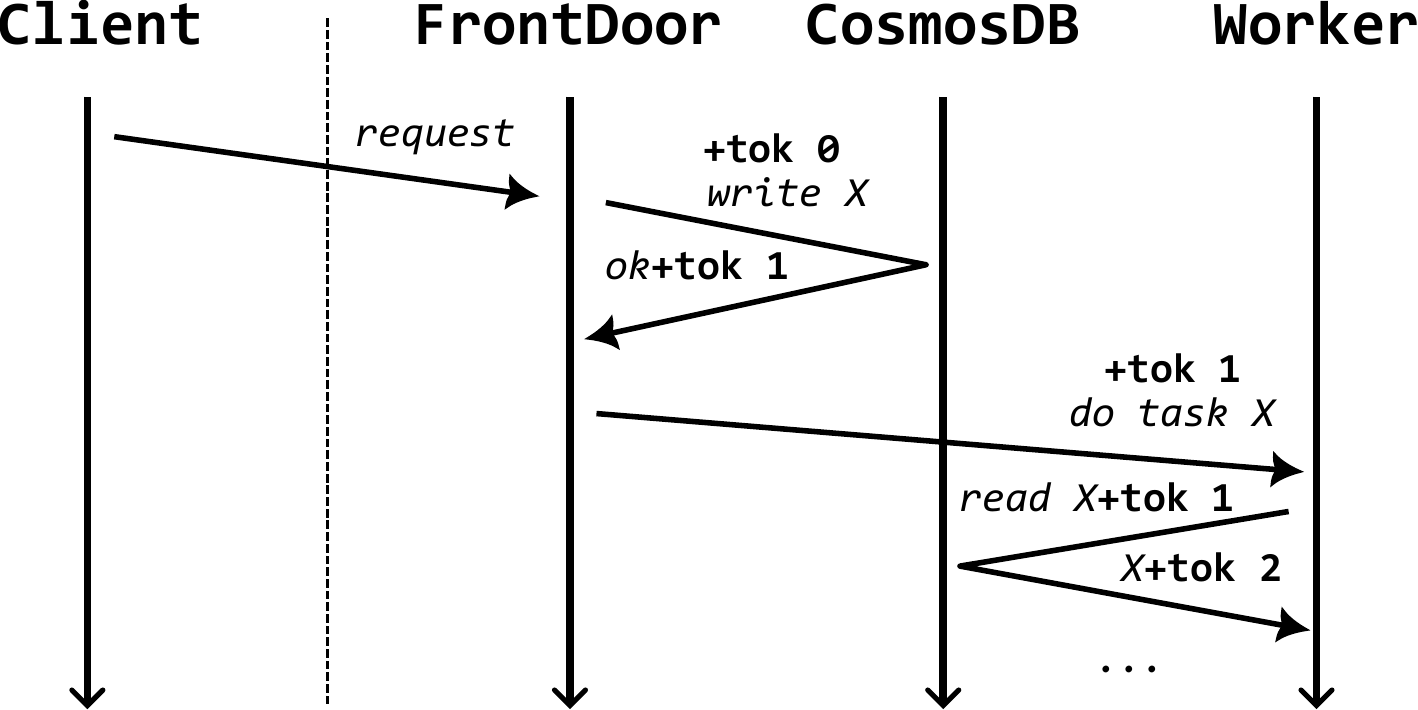}
    \caption{Amended interaction diagram showing a correction of the problem in Figure \ref{fig:sev1-A}.}
    \label{fig:sev1-B}
\end{figure}

Our proposed semantic fix is to pass a session token from \texttt{frontdoor} to \texttt{worker}, and use it as a starting value for \texttt{workerToken} instead of \texttt{CosmosDB!NoSessionToken} when the worker reads from Cosmos DB.
Making this change and re-checking the model, we find that the error no longer occurs.
Figure \ref{fig:sev1-B} illustrates the modified behavior we expect, with versioned $\texttt{+tok}\,0,1\dots$ annotations indicating the process of passing along and keeping up to date a session token \texttt{tok} between \textit{FrontDoor} and \textit{Worker}.

\subsection{Notable Anomalies} \label{scn:notable-anomalies}

By writing our specification, we found multiple anomalous behaviors that we suspected to be specification bugs.
However, in each case it has been confirmed that these represent real behaviors of which Cosmos DB is capable.
These behaviors are not explicitly mentioned in the documentation, and we discovered them purely by model checking, manually examining the semantics of our specification, and discussing our results with author 2, a Cosmos DB expert.

\subsubsection{Dirty Reads} \label{scn:dirty-reads}

Strongly consistent writes, which we expect to be linearizable, are only linearizable in relation to strongly consistent reads.
Other read consistency levels allow dirty reads, which do not follow linearizability.
Since write operations are not atomic, reads with a consistency level other than strong may see incomplete writes, because they are able to see non-durable writes in general.

Following a more implementation-focused analogy, session and eventual consistency reads are only served by one replica in a region.
Each replica immediately begins serving writes it receives without waiting for the writes to fully replicate, so if a read request reaches a replica holding an unreplicated write, then that read can witness an in-progress strongly consistent write operation.
A similar scenario is possible for bounded staleness reads under strong consistency, where a write might be replicated to one region but not a global majority; a bounded staleness read can likewise be served by a region that stores an in-progress strongly consistent write.

The non-atomicity of strongly consistent writes is counter-intuitive and not well-known.
Previous drafts of the specification we present did not include this feature, until we described to author 2 that our formulation effectively assigned concurrency barrier semantics to strong consistency writes, which is incorrect.
The resulting explanation of the true guarantees offered by strong consistency writes inspired the current version of our specification.

\subsubsection{Durable Failed Writes} \label{scn:durable-failed-writes}

Clients may also read the values written by failed writes.
Our specification does not remove or invalidate log entries when a client might observe a write failure, because the write may still succeed after that point.
When in doubt, Cosmos DB will still complete a write operation even if its notification does not reach the requesting client.

While this anomaly is well known, it can be counter-intuitive to developers.
Future documentation may benefit from discussing this possibility, as well as the particular trade-off made by Cosmos DB.

\subsubsection{Bounded Staleness Reads are Weaker Under Strongly Consistent Writes} \label{scn:bounded-staleness-weaker}

Because the guarantees underlying bounded staleness are enforced at write time, performing a bounded staleness read while Cosmos DB is configured for strongly consistent writes does not actually guarantee the same set of bounds as when bounded staleness writes are configured.
Under strong consistency, there is no bound on the number of in-progress write requests.
As a result, bounded staleness reads are not subject to any bounds either, and will return either the same result as a strongly consistent read, or a dirty read from an in-progress strongly consistent write.

We do not expect this more obscure anomaly to cause problems for developers in practice, but it is important to keep note of it in documentation and future design discussions.

\section{Discussion}

The results of our work specifying Cosmos DB shows that a minimal, purely client-facing model of a sufficiently complex distributed system has many uses in practice.
Our specification effort enabled us to suggest several improvements to Cosmos DB's public-facing documentation, as well as to precisely diagnose the root cause of a high-impact outage within Azure Cloud.

Our outcome is a useful intersection between focusing on implementation correctness and focusing on the purely theoretical properties of an abstract \emph{kind of system}.
By keeping our specification at the interface level, we were able to successfully avoid the complexity of Cosmos DB's low-level implementation semantics, while still producing useful practical insights into the behavior of the system we studied.
Furthermore, because of our level of abstraction, we never needed to materially interact with the Cosmos DB implementation itself, beyond person to person discussions with author 2.

Our lack of interaction with Cosmos DB's implementation is a double-edged sword, in that it is possible that some error in our specification has escaped the notice of those reviewing it.
If that has happened, any such error could only be discovered by manual review.
In theory, techniques such as trace validation~\cite{tlatraceverification2005} are relevant, but they are impractical due to the size and complexity of Cosmos DB's implementation.
While this is a fundamental limitation of our approach, our \tlaplus\ model is compact enough that isolating and fixing any error is not difficult: our core specification is only $390$ lines long, including comments and whitespace.
For example, once we were informed dirty reads should be possible, it took us only 2-3 days to rewrite our model's write semantics from fully atomic to the current two-step version, then adapt and re-verify any affected correctness conditions.

This is why we believe that, despite the lack of automated linkage between our model and Cosmos DB's implementation, it is practical to keep the model up to date in the face of any significant design or implementation changes to Cosmos DB.
In fact, analysis of what effect a design change would have on client-observable behavior would likely be beneficial to the discussion of that design change.
Additionally, \tlaplus\ can be used to explore refinement relationships between our client-level specification and other internal implementation-level specifications, such as those currently in use by the Cosmos DB development team.

Outside of Cosmos DB's documentation, our work can be used to precisely model individual interactions with Cosmos DB, which opens up possibilities for using formal methods in the design process of systems where it was previously infeasible due to the need for a re-usable model of Cosmos DB.
This is made possible by our focus on specifying Cosmos DB's interface, since implementation-level models will often be too complex, or have a larger state space than is viable for model checking, to easily be used as components of other models.

\section{Related Work}

There exist multiple perspectives on studying the observable behavior of distributed key-value stores: abstract formal reasoning, formal methods operating on both specifications and implementations, and client-level testing tools.

Formally, database consistency properties have been well studied in the abstract~\cite{serializabilityEC2017,linearizability1990}.
In particular, \cite{seeingIsBelieving2017}'s focus on client-observable system states partially inspired our modeling strategy for Cosmos DB.

In formal methods, efforts are ongoing to specify and verify the correctness of distributed system implementations.
Verifying that an implementation satisfies a given specification can be a powerful tool, but it often requires that the implementation has a specific structure, often requiring verification to be part of the development process from the beginning~\cite{shardstore2021,everest2017}, or at least deeply integrated into the development process~\cite{amazons2n2018}.
The adoption cost of such techniques may prove prohibitive for existing large, unverified codebases.

Tools to explore possible behaviors of an unmodified implementation have been successfully developed~\cite{yang2009modist,chaosstudio2022}, but these tools focus on exposing implementation bugs rather than studying the set of valid client-observable behaviors.

Client-level testing tools also exist~\cite{coyote2021,cosmossimulator2022}, but this work focuses on more general-purpose anomalies, or relies on user-provided definitions for dependencies like databases.
Database semantics, especially quirks of a specific implementation, are hard to define and reason about.
Mock implementations and simulation modes for complex database services cannot be built as an afterthought.
Our work provides a well-reasoned starting point for building any client-level testing tools specialized to Cosmos DB and its anomalies.

MonkeyDB~\cite{monkeydb2021} provides a general-purpose definition of database consistency semantics, which it uses to simulate client code interactions with databases.
We believe our approach is complementary to this kind of more general-purpose simulator, in that our implementation-specific specification offers a different set of semantics to simulate, potentially including quirks that are unique to Cosmos DB.

Elle~\cite{elle2020} automatically validates database consistency guarantees by analyzing the outcomes of synthetic query sequences.
It may be useful in both exploring the actual semantics of a black-box database implementation, and in data consistency bug-finding.

\section{Conclusion}

We have presented what can best be described as \emph{the lightest-weight useful specification of Azure Cosmos DB's semantics in \tlaplus}.
Despite its structural simplicity, our model covers all 5 advertised data consistency levels available to clients.
It represents behaviors with arbitrary configurations of regions and replicas, including arbitrarily complex scenarios involving delayed replication, server and region failure, and otherwise data loss.

Our new specification has been validated by a combination of model checking, refinement with existing incomplete specifications, and expert review.
While we are now confident in our model's correctness, should any bugs be found in it, our model is also small enough that fixing them would not require inordinate amounts of work.

We have used our model to predict multiple under-documented anomalous behaviors of Cosmos DB, and to raise two now-addressed issues with the service's publicly-available documentation.
We have also used our model to elaborate on the root cause of a high-impact outage within Azure Cloud, successfully producing an abstract explanation for the underlying series of events.

In the future, we expect our model to be be usable by the Cosmos DB development team to reason about their service's client-facing behavior, in conjunction with their own implementation-level \tlaplus\ specifications via refinement.
Beyond benefits to Cosmos DB specifically, our compact specification can also be used to specify systems dependent on Cosmos DB, growing the set of systems for which formal verification is viable.

Our results show the value of using formal verification in industry, even without any interaction with the target system's implementation at all.
The benefits in terms of understanding and documenting a system's expected behavior are still significant for end-users and developers.

\section*{Acknowledgment}

We thank the authors of existing specifications of Cosmos DB, whose work we used as a starting point: Murat Demirbas, Ailidani Ailijiang, and Dharma Shukla.
We also thank Microsoft engineer Ben Pannell for helping us understand the cloud incident discussed in Section \ref{scn:sev1}.
We thank Leslie Lamport and Ivan Beschastnikh for their writing feedback.

\bibliographystyle{IEEEtran}
\bibliography{IEEEabrv,paper}






\end{document}